\documentclass[twocolumn,a4paper,nofootinbib,aps,prd,showpacs]{revtex4}
\usepackage{graphics}
\usepackage{wasysym}
\usepackage{url}
\usepackage{comment}
\usepackage{hyperref}
\usepackage{setspace}
\usepackage{bm}

\newcommand{\gw}{gravitational wave }
\newcommand{\gws}{gravitational waves }
\newcommand{\Othresh}{O_{\textrm{\scriptsize{thresh}}}}
\newcommand{\subgw}{_{\textrm{\scriptsize{GW}}}}
\newcommand{\ee}[1]{\!\times\!10^{#1}}

\begin{document}

\title{An Evidence Based Search Method For Gravitational Waves From Neutron Star Ring-downs}
\author{James Clark}
\email{jclark@astro.gla.ac.uk}
\author{Ik Siong Heng}
\email{siong@astro.gla.ac.uk}
\author{Matthew Pitkin}
\email{matthew@astro.gla.ac.uk}
\author{Graham Woan}
\email{graham@astro.gla.ac.uk}
\affiliation{Department of Physics and Astronomy, Kelvin Building, Glasgow G12 8QQ, UK}

\date{\today}

\singlespacing

\begin{abstract}
The excitation of quadrupolar quasi-normal modes in a neutron star
leads to the emission of a short, distinctive, burst of
gravitational radiation in the form of a decaying sinusoid or
`ring-down'. We present a Bayesian analysis method which
incorporates relevant prior information about the source and known
instrumental artifacts to conduct a robust search for the
gravitational wave emission associated with pulsar glitches and soft
$\gamma$-ray repeater flares.  Instrumental transients are modelled
as sine-Gaussian and their evidence, or marginal likelihood, is
compared with that of Gaussian white noise and ring-downs via the
`odds-ratio'.  Tests using simulated data with a noise spectral
density similar to the LIGO interferometer around $1$ kHz yield
$50\%$ detection efficiency and $1\%$ false alarm probability for
ring-down signals with signal-to-noise ratio $\rho=5.2$.  For a
source at $15$ kpc this requires an energy of $1.3\times
10^{-5}M_{\astrosun}c^2$ to be emitted as gravitational waves.

\end{abstract}

\preprint{LIGO-P070008-01-Z}

\maketitle

\section{Introduction}\label{sec:intro}

A possible mechanism for the emission of \gws from neutron stars is
the excitation of non-radial quasi-normal modes (QNMs)
\cite{thorne:1969}.  This excitation could be caused by the
disruption associated with pulsar glitches
\cite{andersson_fingerprints} or from flaring activity in soft
$\gamma$-ray repeaters \cite{pacheco}.

The frequencies and damping times of the QNMs depend strongly on the
neutron star equation of state (EOS) and for the more dominant
\emph{f}-modes these are thought to lie somewhere in the region of
$1$ to $4$ kHz and $50$ to $500$\,ms, respectively. Andersson
\& Kokkotas \cite{andersson} have shown how the mass and radius of a
neutron star may be constrained by \gw observations of the QNM
frequencies and decay times.  Conversely, if the EOS of the neutron
star were known with some precision, it would be possible to compute
the expected decay times and frequencies of the QNMs. This would
provide a well constrained waveform model for aiding the
identification of a potential \gw signal following a neutron star
ring-down.  However, the behaviour of matter at the densities found
in neutron stars is not well understood and there exist many
different models for the neutron star EOS.  It is, therefore,
necessary to develop techniques which are robust to the
uncertainties in these models.

While the \gw emission from neutron star QNMs is expected to be weak
(inducing typical strain amplitudes of $\sim 10^{-24}$), their
detection is further hampered by the presence of instrumental
glitches that can closely resemble short-duration \gw signals.

In this work, we demonstrate the feasibility of applying Bayesian
inference to the robust detection of neutron star \gw ring-downs
through a process of model selection. We highlight how the
methodology may be extended to include a more realistic range of
glitches and developed into a multi-detector search.


\section{Perturbed neutron stars \& asteroseismology}\label{sec:neutron_stars}
When the solid crust of a neutron star is severely disrupted or
cracked, some of the stored elastic energy is channeled into the
oscillatory modes of the star. Quadrupolar excitations will then be
strongly damped by \gw emission \cite{thorne:1969}.   The different
modes may be labelled according to the spherical harmonic indices
$l$ and $m$ which describe the angular dependence and number of
nodes.

The fundamental fluid mode, or $f$-mode (as first shown by Kelvin
for the case of a non-rotating, uniform density star) has angular
frequency
\begin{equation}
\omega_f^2 = \frac{2l(l-1)}{2l+1} \frac{GM}{R^3},
\end{equation}
where $M$ and $R$ are the stellar mass and radius, respectively.
This is also a reasonable estimate for more realistic equations of
state \cite{andersson_review} and we see that, for  non-radial modes
with $l=2$, the $f$-mode pulsations have frequencies $\sim 2$ kHz
taking the fiducial values $M = 1.4M_{\astrosun}$ and $R = 10$ km.
Other modes, such as the pressure ($p$) and space-time ($w$) modes,
have considerably higher frequencies than this. Gravitational wave
interferometers like GEO600 \cite{GEO}, LIGO \cite{LIGO} and VIRGO
\cite{VIRGO} are more sensitive at lower
frequencies making the $f$-mode the most favourable for a \gw
search. The ring-down timescale, $\tau_f$, is given by the ratio of
the oscillation energy to the total power emitted as \gws
\cite{andersson}. This yields $\tau_f \sim R(R/M)^3$, where $M$ and
$R$ are the mass and radius of the neutron star, respectively.

So, given some EOS, which defines the ratio $M/R$,  it is possible
to calculate the exact damping times and frequencies of the QNMs.
Andersson \& Kokkotas \cite{andersson} do precisely this for a
variety of equations of state and establish empirical relations for
the $f$-mode damping time and frequency.  On the other hand, we can
consider the inverse problem and use \gw observations of the
$f$-mode frequency and damping time to constrain the neutron star
EOS. Indeed, there has already been an attempt to use the
electromagnetically observed frequencies of potential torsional mode
oscillations in two SGRs to constrain their equations of state
\cite{Samuelsson:2006}.

\begin{center}
\begin{figure}[]
    \includegraphics{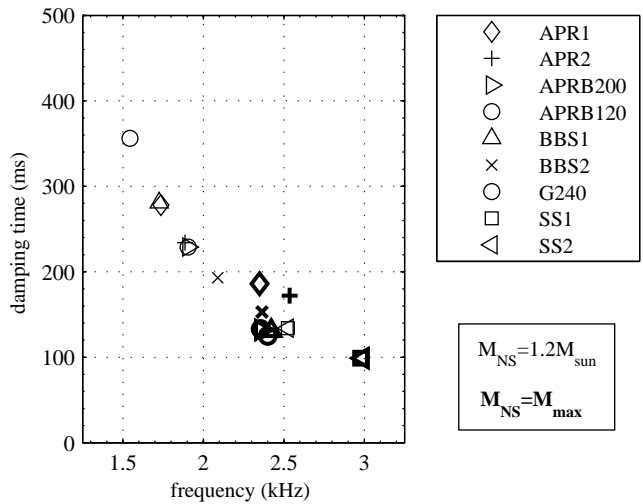}
    \caption[\emph{f}-mode frequencies \& damping times for various NS EOS's]{The \emph{f}-mode
    frequencies \& damping times for the variety of different equations of state (indicated by
    symbol shape) and the extremal neutron star masses (indicated by symbol
weight) considered in \cite{benhar}.}
    \label{fig:model_eigenvalues}
\end{figure}
\end{center}

Here, our interest is in using what little \emph{is} known about the
neutron  star equation of state to inform us with regards to
sensible \gw waveforms to search for.
Fig.~\ref{fig:model_eigenvalues} shows the results of calculations
by Benhar {\it et al} \cite{benhar} of the $f$-mode frequency and
damping time, using several realistic EOSs and the extremal neutron star masses. 
$\rm M_{\rm max}$ refers to the maximum neutron star mass allowed by the EOS.  The reader is
directed to \cite{benhar} and the references
therein for descriptions of the different equations of state.

Examining Fig.~\ref{fig:model_eigenvalues} we see that the equations
of state considered in \cite{benhar} yield typical $f$-mode
frequencies of $\nu_f \sim 1.5 - 3$ kHz and damping timescales of
$\tau_f \sim 50 - 400$\,ms.  Later, we use these ranges to set
sensible limits of our priors for each parameter.

\subsection*{Gravitational wave emission}
Following Thorne \cite{thorne:1969}, we model the gravitational wave
strain amplitude at the Earth from the $n$-th QNM as:
\begin{equation}\label{eqn:waveform}
\begin{tabular}{cl}
 $h(t) = \left\{ \begin{tabular}{cl}
 $ h_0\sin\left[\omega_n \left(t-t_0\right) \right]e^{\left[-\frac{(t-t_0)}{\tau_n }\right]}$
&for $
t\geq t_0$ \\ \\
 $0 $ &otherwise, $  $ \\
\end{tabular} \right.$\\
\end{tabular}
\end{equation}
where $h_0$, $\omega_n$ and $\tau_n$ are the initial amplitude,
angular frequency and characteristic damping time of the signal, and
$t_0$ is the start time of the signal. We take $n=0$ to represent
the $f$-mode.

Furthermore, $h_0$, $\omega_0$ and $\tau$ are related to the total
\gw energy via the following relation \cite{pacheco},

\begin{equation}\label{eqn:GWenergy}
E_{\textrm{\scriptsize{GW}}} = \frac{c^3 D^2}{4G}\left(h_0 \omega_0
\tau^{\frac{1}{2}}\right)^2,
\end{equation}

where $D$ is the distance to the source.  We can, therefore, write
down an expression for the the expected initial amplitude for
$f$-mode ring-downs using fiducial parameter values:
\begin{widetext}
\begin{equation}\label{eq:canonical}
h_0 \sim 1.6\times10^{-24}
\left(\frac{E_{\textrm{\scriptsize{GW}}}}{10^{-11}M_{\astrosun}c^2}\right)^{1/2}
\left(\frac{\tau}{200 \ \textrm{ms}}\right)^{-1/2} \left(\frac{\nu_0}{2 \ \textrm{kHz}}\right)^{-1}
\left(\frac{D}{15 \
\textrm{kpc}}\right)^{-1}.
\end{equation}
\end{widetext}

\subsection*{Search triggers \& QNM excitation}
There are several ways to generate stellar pulsations including
close encounters with orbital companions and accretion (e.g., of
comets), the ringing of proto-neutron stars following supernovae,
soft $\gamma$-ray repeater (SGR) flaring activity and crustal
disruption due to pulsar glitches.

Our aim is to search for \gws from the latter two mechanisms.  This
is because both pulsar glitches and the SGRs provide an observable
electromagnetic counterpart (which may be used to `trigger' the
search) and occur with a frequency making them practical for a
triggered search. The most prolifically glitching pulsar,
PSR\,J0537-6910, has a glitch rate of $\sim 4\,{\rm year}^{-1}$
\cite{Middleditch:2006}, whilst other regular glitchers like the
Crab pulsar and Vela have rates of order $\sim 0.5\,{\rm year}^{-1}$
\cite{CrabEphemeris, Dodson:2006}. SGRs are seen to emit bursts at
rates of a few to $10\,{\rm year}^{-1}$ \cite{Nakagawa:2007},
whereas hyperflares are much rarer events occuring maybe once a
decade. Other potential candidates seen to glitch are the anomalous
X-ray pulsars (AXPs).

\subsubsection*{Pulsar glitches}
Pulsar glitches are observed as sudden irregularities in the
rotation rate of pulsars and are characterised by a step increase in
rotation frequency. The characteristics of glitches vary between
pulsars. For example, some show step changes in the spin-down rate
and others exponential recoveries to pre-glitch parameters. The
exact characteristics give clues to the underlying mechanisms
causing the glitch.

The mechanisms responsible for pulsar glitches are still unclear,
but there are two main candidates to explain the underlying process
and some of the differences between glitches. For older pulsars it
seems glitches are likely caused by a dramatic decoupling between
the star's solid crust and superfluid interior \cite{Cheng:1988}.
Glitches in the young Crab pulsar are thought to be associated with
a reconfiguration of the solid crust \cite{Franco:2000}: the
spin-down reduces the centrifugal force and the crust reaches
breaking strain.  The ensuing relaxation of ellipticity will cause a
sudden change in the moment of inertia, producing the observed
glitch, and the crustal rupture will set up a starquake, hopefully
causing $f$-mode excitations. The reality of the mechanism is likely
to be very complex and may be a combination of the two.

For the two different glitch models the amount of energy released can be estimated in
different ways as shown in van~Riper {\it et al.} (1991) \cite{vanRiper:1991}. They assume all the
energy released goes into heating the star, whereas we will make the assumption that it goes into
exciting quasi-normal modes. For the angular momentum exchange model (thought to be the most
probable explanation for the Vela pulsar glitches) the amount of energy released depends on the
angular momentum exchanged between the superfluid interior and crust $\Delta{}J \sim I\Delta\Omega$,
where $\Delta\Omega$ is the angular frequency change from the glitch. The energy change is then
$\Delta{}E = \Delta{}J\Omega_{\rm lag}$, where $\Omega_{\rm lag}$ is the lag frequency between the
superfluid and crust, with an estimated range of values of 1-100\,rad\,${\rm s}^{-1}$ (or possibly
$\lesssim 0.1$\,rad\,${\rm s}^{-1}$) \cite{vanRiper:1991}. For the largest Vela pulsar glitch, with
a fractional frequency change of $\Delta\Omega/\Omega = 3.1\ee{-6}$ \cite{Dodson:2002}, this gives a
$\Delta{}J \sim 2\ee{34}$\,J giving an energy release of $\Delta{}E \sim 10^{-13} - 10^{-11}\,{\rm
M}_{\odot}c^2$ for the range $\Omega_{\rm lag} \approx$ 1-100\,rad\,${\rm s}^{-1}$.

For starquake driven glitches the energy released is given in Ref.~\cite{vanRiper:1991} as
$\Delta{}E \approx \mu{}V_{\rm crust}\epsilon_{\rm max}\epsilon_{\rm quake}$, where $\epsilon_{\rm
quake} = \Delta\Omega/\Omega$ is equivalent to the relative change in moment of inertia, $\mu$ is
the mean shear modulus of the star, $V_{\rm crust}$ is the volume of the crust (where $\mu{}V_{\rm
crust} \sim 10^{41}$\,J), and $\epsilon_{\rm max}$ is the maximum deformation from equilibrium the
crust can withstand without breaking (given in Ref.~\cite{vanRiper:1991} as $\epsilon_{\rm max}
\lesssim 10^{-2}$ although this could vary somewhat). Assuming this $\epsilon_{\rm max}$ and taking
the largest Crab pulsar glitch, where $\Delta\Omega/\Omega \sim 8\ee{-8}$ \cite{Lyne:1993}, we get
an energy release of $\Delta{}E \sim 5\ee{-16}\,{\rm M}_{\odot}c^2$. If the starquake mechanism can
provide similar fractional frequency changes to a neutron star to those seen in the Vela pulsar
during glitches, then this mechanism could still be a valuable potential source.

\subsubsection*{Soft $\gamma$-ray repeater flares}
Soft $\gamma$-ray repeaters are high energy transient sources with
typical photon energies of $10-30$ keV and similar burst
characteristics from one event to the next, although they are also
seen as quiescent X-ray sources. These objects are identified as
highly magnetised ($B \approx 10^{14}$ Gauss) neutron stars, or
`magnetars'. They are occasionally seen to emit giant flares, or
hyper-flares, which have thousands of times the luminosity, and
harder spectra, than the regular bursts. The hyperflares are thought
to occur when magnetic field becomes twisted and causes a
catastrophic reconnection, inducing tectonic activity. The field
annihilation/reconnection in seismic faults is responsible for the
observed $\gamma$-ray emission and, again, we expect the crustal
disruption to excite the QNMs of the magnetar \cite{Hurley:2005}.

\section{Bayesian Model Selection}
Before moving on to describe the search in detail, we outline some
ideas behind Bayesian model selection.

Bayes' theorem describes how to assign a posterior probability to
some parameterised model $M_i$, given a set of data or observations $D$ and some
background information $I$ which determines the hypothesis space
$\{M_i\}$:
\begin{equation}\label{eqn:bayes}
    p(M_i|D,I) = \frac{p(M_i|I)p(D|M_i,I)}{p(D|I)}.
\end{equation}
Here $p(D|M_i,I)$ is the \emph{marginal likelihood} or \emph{evidence} and
represents the influence of the data on our belief in $M_i$\footnote{The name
\emph{marginal likelihood} reflects the fact that we have marginalised over the parameter
values that are associated with $M_i$.}; $p(M_i|I)$ is the
\emph{prior} probability of model $M_i$ and describes our state of
belief in $M_i$, preceding examination of the data $D$; $p(D|I)$ ensures that the
posterior is correctly normalised.

Our desire here is to detect a gravitational wave of a known shape
(see Eq.~\ref{eqn:waveform}) but with unknown parameters within some
range in noisy data. The obvious question we might ask is, `does
this data contain a ring-down gravitational wave?'.  We denote the
data by $D$ and the gravitational waveform model by $M_{\textrm{\scriptsize GW}}$.

The posterior probability $p(M_i|D,I)$
then tells us the degree of belief to assign to the model
$M_{i}$.  To properly normalise the posterior,
however, we must marginalise over the entire hypothesis space:
\begin{equation}\label{eqn:evidence_sum}
    p(D|I) = \sum_i p(M_i|I)p(D|M_i,I).
\end{equation}

If the system is sufficiently well understood, it is possible to
enumerate all possible models $M_i$ and the posterior probability of
any one model can be calculated directly from Bayes theorem.
However, this cannot always be done and it often makes more sense to
evaluate the probability of one model relative to another.  Such
comparisons are performed via the \emph{odds ratio}:
\begin{equation}\label{eqn:posterior_odds}
    O_{12} = \frac{p(M_1|D,I)}{p(M_2|D,I)}.
\end{equation}
For a gravitational wave search we might choose $M_1$ to be the proposition that the
noisy data contains a gravitational wave, and $M_2$ to be the proposition that the data
only contains detector noise.

Substituting the right hand side of Bayes' theorem for the
posteriors in Eq.~\ref{eqn:posterior_odds}, we see that the
normalisation term $p(D|I)$ drops out and we are left with
\begin{equation}\label{eqn:full_odds}
    O_{12} = \frac{p(M_1|I)}{p(M_2|I)} \frac{p(D|M_1,I)}{p(D|M_2,I)}.
\end{equation}
The first term, the prior odds, is the ratio of the prior
probabilities for each model. Typically, we assume complete naivete
and set the prior odds equal to unity.  The second term is the ratio
of the evidences from each model and is called the Bayes factor.
Clearly a large value of the Bayes factor indicates a strong
preference for $M_1$.

The evidence is computed by integrating the likelihood $p(D|{\bm \theta}, M_i, I)$
over all model parameters ${\bm \theta}$ and weighting by the prior on those parameters,
$p({\bm \theta}|I)$, leading to the alternative name
`marginal likelihood':
\begin{equation}\label{eqn:model_likelihood}
    p(D|M_i,I) = \int_{{\bm \theta}}
    p({{\bm \theta}}|M_i,I) p(D|{{\bm \theta}},M_i,I)
{\rm d}{{\bm \theta}}.
\end{equation}
So given some competing models $M_1$ and $M_2$, we can evaluate the
evidences $p(D|M_1,I)$, $p(D|M_2,I)$ and assuming prior odds of
unity, use the odds ratio to decide which is most likely, given the
data $D$.

\subsection{Application \& choice of models}
Here, `model' shall refer to a \emph{class} of descriptions for the
data.  An example is `the data contains a ring-down gravitational
wave signal in addition to noise'.  Note that we have defined the
generic shape of the data (a noisy ring-down) but not any parameter
values.  A particular signal with a specified set of parameter
values is called a `template'.  In this way, a model defines a set
of templates with parameter values determined by the priors in that
model. Additionally, when we talk about the `evidence for the
model', we are referring to the marginal likelihood for that model,
i.e., $p(D|M,I)$.  The total evidence of the hypothesis space $p(D|I)$ is
eliminated through the use of the odds ratio.

\subsubsection*{$M_1$: Ring-down waveform \& Gaussian white noise}
The expression for the ring-down waveform $h(t)$ is given by
Eq.~\ref{eqn:waveform}.  We assume that the noise $n(t)$ is white
and Gaussian over a sufficiently broad band and that the data stream is given by
\begin{equation}
    d(t) = h(t) + n(t),
\end{equation}
where the noise $n(t)$ has zero mean and variance $\sigma_n^2$.
However, to simplify data conditioning, we work with the power
spectral density,
\begin{equation}\label{eqn:data}
    D(\omega) = |\tilde{h}(\omega)|^2 + |\tilde{n}(\omega)|^2 +
2|\tilde{h(\omega)}\tilde{n(\omega)}|,
\end{equation}
where $\tilde{h}(\omega)$ is the Fourier transform of $h(t)$,
\begin{equation}\label{eqn:h_of_omega}
    |\tilde{h}(\omega)| = \frac{h_0 \tau}{\sqrt{1+(\omega - \omega_0)^2\tau^2}},
\end{equation}
so that our parameter space is given by ${\bm \theta} = \{h_0,
\omega_0, \tau\}$. Notice that working with the power spectral
density has the effect of pre-marginalising over the start time of the
signal, $t_0$ with a uniform prior. The power spectral density is estimated from fast
Fourier transforms (FFTs) of consecutive segments of the time
series.  This yields a spectrogram with time bins indexed by $i$ and
frequency bins indexed by $j$.

The likelihood of obtaining power $D_{ij}$ in the $i$-th time bin at
the $j$-th frequency (i.e., the $i,j$-th pixel), given a template
with signal power $S_{ij}=|\tilde{h}(\omega_j)|^2$ is a non-central
$\chi^2$ distribution with two degrees of freedom and non-centrality
parameter equal to the power from the template, i.e.
\begin{widetext}
\begin{equation}\label{eqn:M1_likelihood}
    p(D_{ij}|S_{ij},M_1,I) = \frac{1}{2\sigma^2_{ij}}
    \exp\left\{-\frac{D_{ij} +
    S_{ij}}{2\sigma^2_{ij}}\right\}I_0\left(\frac{\sqrt{D_{ij}S_{ij}}}{\sigma_{ij}^2}\right),
\end{equation}
\end{widetext}
where $\sigma^2_{ij}$ is the variance of the Fourier components in
that pixel and $I_0$ is the zeroth order modified Bessel function of
the first kind.  If the data is Gaussian and white, the power
spectral density can be normalised such that the Fourier components
follow Gaussian distributions with zero mean and unity variance
(i.e., $\sigma^2_{ij} = 1$).

To calculate a single odds ratio in each time bin, we require the
joint probability across frequencies,
\begin{equation}\label{eqn:jointL}
    p(\{D\}|{\bm \theta},M_1,I) = \prod_j
p(D_{j}|{\bm \theta},M_1,I),
\end{equation}
where we have dropped the time bin index $i$ for notational
convenience. Finally, we adopt independent, uniform priors on the
parameters ${\bm \theta} = \{h_0,\omega_0,\tau\}$ with ranges on
$\omega_0$ and $\tau$ based loosely on the expected values in the
literature.  For the amplitude $h_0$, however, there is little to be
gained by restricting the prior range and the prior on $h_0$ is
taken to run from zero to some arbitrarily high value.  The joint
prior is then
\begin{equation}\label{eqn:joint_prior}
    p({\bm \theta}|M_1,I) = p(h_0|M_1,I)p(\omega_0|M_1,I)p(\tau|M_1,I)
\end{equation}
where
\begin{equation}
\begin{tabular}{cl}
 $p(\theta|M_1,I) = \left\{ \begin{tabular}{cl}
 $\frac{1}{\theta_{\rm max}-\theta_{\rm min}} $ & for $ \theta_{\rm min} \leq \theta \leq
\theta_{\rm max}$ \\ \\
 $ 0 $ &otherwise, $  $ \\
\end{tabular} \right.$\\
\end{tabular}
\end{equation}
and $\theta$ is any one of the parameters in model $1$.

\subsubsection*{$M_2$:  Gaussian white noise only}
$M_2$ is our null detection hypothesis: the data is modelled as
white Gaussian noise, without any gravitational wave signal, so that
$d(t) = n(t)$. Again, we work in terms of the power in each
spectrogram pixel $D_{ij}$. If the spectrogram has been normalised
such that the individual Fourier components are normally distributed with
mean zero and variance of unity, we know that the power in the $j$-th frequency bin
follows a central $\chi^2$ distribution with two degrees of freedom,
\begin{equation}\label{eqn:M2_likelihood}
    p(D_{ij}|M_2,I) = \frac{1}{2\sigma^2_j}\exp\left\{-\frac{D_{ij}}{2\sigma^2_j}\right\},
\end{equation}
where $\sigma^2_j$ is the variance of the Fourier components in the $j$-th frequency bin. 
If it is possible to estimate the variance in each frequency bin (from an
off-source piece of data, for example), then the result is also valid for coloured noise.

Notice that we have arrived at the evidence from $M_2$
without making any mention of parameterisation, priors or marginalisation.
This can also be derived in a purely Bayesian context by considering
the likelihood of a pixel power $D_{ij}$, given a template power
$S_{ij}$ (Eq.~\ref{eqn:M1_likelihood}).  In the case of a model
where there is no contribution to the power from a gravitational
wave, we know \emph{a priori} that $S_{ij} = 0$.  That is to say
\begin{equation}\label{eqn:M2_powerprior}
    p(S_{ij}|M_2,I) = \delta(S_{ij}),
\end{equation}
where $\delta$ is the Dirac delta function. If we now marginalise
the likelihood given by Eq.~\ref{eqn:M1_likelihood} over power using
this prior, we arrive at Eq.~\ref{eqn:M2_likelihood}.

\subsubsection*{Thresholding $O_{12}$}
$O_{12}$ is the ratio of the posterior probabilities for each model,
so it might seem sensible to choose $O_{12}>1$ to indicate a
preference for $M_1$ over $M_2$.  While this is true, such a
threshold for \gw detection neglects the role of our prior odds and
the need for an acceptable false alarm rate.

By setting the prior odds equal to unity, we are saying we believe
\emph{a priori} that both models are equally probable.  Even if we
truly were that ignorant, the influence of the data through the
Bayes factor may cause the odds to fluctuate around some
mean value away from unity.  Instead, we search for excesses from the
mean `off-source' (zero-signal) odds to indicate a preference for
our \gw model. Alternatively, it would be straightforward to
estimate the prior odds using an off-source sample.  The value of
the prior odds could then be chosen such that an odds ratio of unity
corresponds to a false detection probability of 0.5. Ultimately, the
odds threshold, denoted $O_{\textrm{\scriptsize thresh}}$, can be
set according to the results of large numbers of trials and so the
overall normalisation is relatively unimportant.

Finally, the magnitudes of the variations of the odds ratios mean that
it is most natural and convenient to work with the base 10 logarithm of
the odds ratio.  For clarity, $\log_{10}  O_{12}$ refers to the base 10 
logarithm of the odds ratio between models $M_1$ and $M_2$.

\subsection{Algorithm \& example}
We now consider an example using data synthesised in
Matlab\footnote{\url{http://www.mathworks.com/products/matlab}} to
illustrate the above principles. First, the outline of the algorithm
is as follows:
\begin{enumerate}
\item   Estimate the variance of the noise using some stretch of data away from the time
of an expected \gw signal (i.e., \emph{off-source}).
\item   Construct the power spectral density of discrete time segments of data centred
around the
expected \gw signal (i.e., \emph{on-source}) to create a time-frequency map of power (spectrogram).
\item   Normalise the power spectral density so that, in the
absence of a signal in the data, the power in a given frequency bin follows a central
$\chi^2$ distribution with
two degrees of freedom.
\item   Compute the evidences $p(D|M_1,I)$, \ $p(D|M_2,I)$ for each model in each
spectrogram time bin.
\item   Assuming prior odds of unity, evaluate the odds ratio $O_{12}$ in each time bin.  An excess
in the odds ratio indicates a preference for $M_1$ and, therefore, a potential detection.
\end{enumerate}
To characterise the signals used to test the algorithm, we define
the signal-to-noise ratio as
\begin{equation}\label{eqn:SNR}
    \rho^2 = 2\int_{-\infty}^{+\infty} \frac{|h(\nu)|^2}{S(\nu)}d\nu,
\end{equation}
where $S(\nu)$ is the one-sided noise spectral density.  In the case
of Gaussian white noise, $S(\nu)$ is given by
\begin{equation}\label{eqn:nosiePSD}
    S(\nu) = 2\frac{\sigma_n^2}{f_s} \ \forall \ \nu,
\end{equation}
where $\sigma_n^2$ is the variance of the time series data and $f_s$
is the sampling frequency.  It is also useful to define the
root-sum-squared amplitude
\begin{equation}\label{eqn:hrss}
    h_{\textrm{\scriptsize{rss}}} = \left(\int_{-\infty}^{+\infty} |h(t)|^2\,{\rm d}t\right)^{1/2}.
\end{equation}
\begin{center}
\begin{figure}[]
    \includegraphics{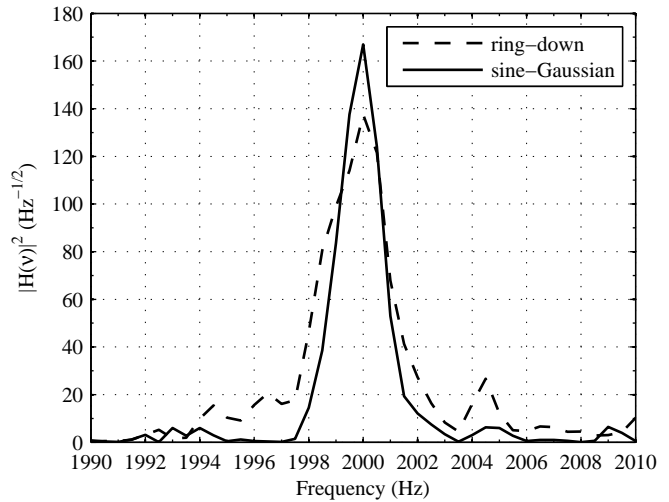}
    \caption[Example power spectra]{The normalised power spectral densities of the
ring-down (dashed line) and sine-Gaussian (solid line) injections.}
    \label{fig:psd}
\end{figure}
\end{center}
To demonstrate the operation of the algorithm, we inject a ring-down
signal into $100$\,s of Gaussian white noise with amplitude spectral
density $10^{-22}$\,Hz$^{-1/2}$.  To investigate the response to a
typical instrumental glitch which closely resembles our target
waveform, we also inject a sine-Gaussian signal of the form
\begin{equation}\label{eqn:SG}
    h(t) = h_0 \sin\left[\omega \left(t-t_0\right)\right] e^{-(t-t_0)^2 / \tau^2},
\end{equation}
where, $h_0$ is the maximum amplitude of the signal, $\omega$ is the
angular frequency and $\tau$ is the decay time.
Table~\ref{table:injs} shows the parameter values used to generate
the injections and Fig.~\ref{fig:psd} shows the power spectral
density of each signal, calculated from the noisy data and
normalised so that the noise follows a central $\chi^2$
distribution. It is the job of the algorithm to detect and
differentiate between these two signals, only producing a candidate
detection when the ring-down is present. The prior ranges used for
each parameter are shown in Table~\ref{table:priors}.
\begin{center}
\begin{table}
\begin{ruledtabular}
\begin{tabular}{l l l c}
Injection & Parameter & & Value \\ \hline
RDI   & Injection time& $t_0$ & $10$s \\
      & Initial amplitude& $h_0 $ & $7.0\times 10^{-21}$ \\
      & Central frequency& $\nu_0$ & 2\,000\,Hz \\
      & Decay time& $\tau$ & 0.21\,s \\
      & SNR& $\rho$ & 20.4 \\ \hline
SGI   & Injection time& $t_0$ & $80$\,s \\
      & Initial amplitude& $h_0$ & $3.72\times10^{-21}$ \\
      & Central frequency& $\nu_0$ & 2\,000\,Hz \\
      & Decay time& $\tau$ & 0.21\,s \\
      & SNR& $\rho$ & 20.4 \\ \hline
\end{tabular}
\end{ruledtabular}
\caption[Example injections]{\label{table:injs}Injected signal parameter values.}
\end{table}
\end{center}

\begin{center}
\begin{figure}[]
    \includegraphics{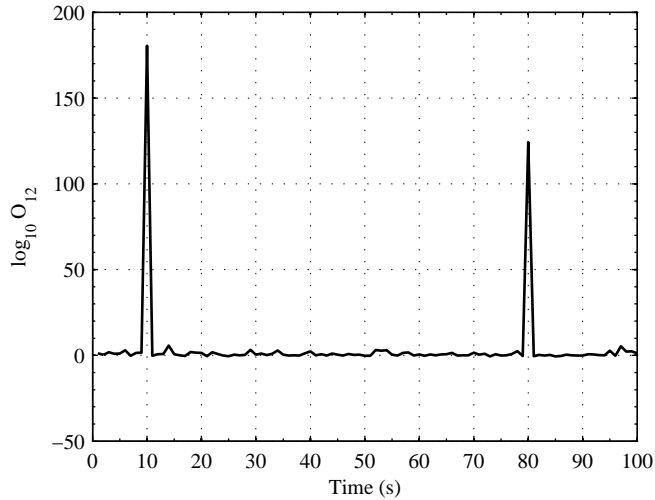}
    \caption[Example ExLods output]{The log-odds in favour of a ring-down
plus noise, versus the possibility that the power is due to noise
alone.  The first peak corresponds to the ring-down injection, the
second to the sine-Gaussian injection.}
    \label{fig:O12}
\end{figure}
\end{center}

\begin{center}
\begin{table}
\begin{ruledtabular}
\begin{tabular}{l c c}
Parameter & Lower Limit & Upper Limit \\ \hline
$h_0 $ & $0$ & $100\times10^{-22}$ \\
$\nu_0$ & $1\,500$\,Hz & $3\,000$\,Hz \\
$\tau$ & $0.05$\,s & $0.5$ s \\
\end{tabular}
\end{ruledtabular}
\caption[Prior ranges]{\label{table:priors}Parameter prior ranges.
The prior distributions are taken as uniform over these ranges.}
\end{table}
\end{center}
Fig.~\ref{fig:O12} shows the odds ratio in each spectrogram time
bin. The ring-down injection at $t=10$ s is strongly detected.
Notice, however, that the sine-Gaussian we have
injected to mimic an unwanted instrumental glitch is also detected, albeit
with an odds ratio lower than that for the ring-down injection. To address 
this issue, we make a straightforward extension to the odds ratio 
to consider multiple hypotheses.

\subsection{Multiple hypothesis extension}
To handle the possibility of sine-Gaussian instrumental glitches, we
rewrite the posterior in the denominator as the sum of the
probability of the noise model $M_2$ and a model for sine-Gaussian
glitches, $M_3$.  This gives
\begin{equation}\label{eqn:null_posterior}
    p(M_{-}|D,I) = p(M_2|D,I) + p(M_3|D,I) + \ldots \ ,
\end{equation}
where $M_{-}$ denotes the proposition that the data does not contain
a ring-down gravitational wave and the ellipsis is to emphasise the
fact that this null-detection hypothesis $M_{-}$ may be further
extended to include additional models for instrumental glitches.
Similarly,
\begin{equation}\label{eqn:pos_posterior}
    p(M_{+}|D,I) = p(M_1|D,I) + \ldots \ ,
\end{equation}
where, again, it is straightforward to include additional signal
models if desired.  The result is that we are left with a comparison
of the probability in favour of a gravitational wave with the
probability of an instrumental glitch or that of the noise model,
thus making maximal use of any knowledge we might have regarding
transient features in the data.

Using $M_{-}$ now as the alternative hypothesis, we obtain a new
odds ratio $O_{123}$:
\begin{equation}\label{eqn:O123}
    O_{123} = \frac{p(M_1|D,I)}{p(M_2|D,I)+p(M_3|D,I)} .
\end{equation}
Assuming the prior on each model is identical, the posterior
probabilities in $O_{123}$ may be replaced by the evidences
$p(D|M_i,I)$ for $i=1,2,3$.

\begin{center}
\begin{figure}[]
    \includegraphics{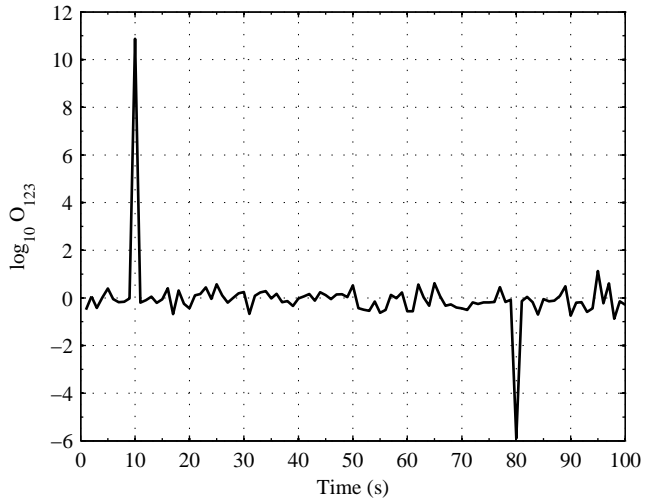}
    \caption[Example ExLods output for $O_{123}$]{The log-odds
in favour of a ring-down plus noise, versus the possibility that the
power is due to noise alone \emph{or} a sine-Gaussian glitch plus noise. The second (sine-Gaussian)
peak seen in Fig.~\ref{fig:O12} is now seen as a dip.}
    \label{fig:O123}
\end{figure}
\end{center}

Fig.~\ref{fig:O123} shows the result of using this new expression
for the odds ratio and the same data from the previous example.
Since the parameterisation for the sine-Gaussian is identical to
that of the ring-down, the priors in table \ref{table:priors} are
also used to evaluate $p(M_3|D,I)$.  There is no fundamental reason
for using the same prior ranges for both models.  Indeed, a
realistic application would most likely have very different priors
for the parameters in different models even if the parameterisation
was the same. Here, the priors in Table~\ref{table:priors} are used
for both $M_1$ and $M_3$ for computational simplicity.

The incorporation of the alternative model eliminates the previous problem of
detecting sine-Gaussian signals with high odds ratios and, in fact, the odds ratio now strongly
prefers the sine-Gaussian model to the ring-down model in the presence of the sine-Gaussian.  
Also note that the size of the peak indicating the
ring-down has diminished quite substantially but is still clearly visible above the
background.  This reduction is due to a non-zero contribution from $p(M_3|D,I)$ in the
denominator of $O_{123}$.  Unless they are mutually exclusive in some way, the inclusion
of additional models in the denominator of the odds ratio will generally increase the
robustness of the search, at the cost of sensitivity.

\section{Performance}
We now investigate the performance of the algorithm by considering
both formulations of the odds ratio, comparing the relative merits
of each.

A `false alarm' is defined as any unwanted transient event which
causes the odds ratio to cross the threshold and generate a
candidate detection.  For white noise, false alarms are caused by spikes in the noise
amplitude.  The false alarm probability
is calculated from the fraction of time bins in a large sample for
which the odds ratio crosses the threshold when there is no
ring-down signal present.

We investigate the response to `off-source' data (i.e., white noise
with no injections) by combining the results of ten 500\,s
spectrograms with a 1\,s time resolution to give 5\,000 off-source
time bins. To evaluate the sensitivity of the search, ring-down
signals of a constant signal-to-noise ratio are injected every other
second into 500 second segments of white noise, synthesised in
Matlab. This is then used to construct a 500\,s spectrogram with
1\,s time-resolution and 0.5\,Hz frequency resolution for each set
of injections, leading to 250 signal injections for each value of
the signal-to-noise ratio.  The fact that there are spectrogram time
bins with no signal injection helps to prevent contamination from
adjacent bins. We vary the signal-to-noise ratio through the value
of $h_0$ only and always compare signals of equal bandwidth and at
the same frequency, with the frequency held constant at $2$\,kHz and the decay time at
$207.5$\,ms. The noise amplitude spectral density is
$10^{-22}$\,Hz$^{-1/2}$, representative of the LIGO noise floor at
these frequencies.

When we make use of the glitch catalogue in the expanded odds ratio
$O_{123}$ the objective is to be robust against unwanted glitches as
well as spurious noise effects.  Therefore, we also define a false
alarm probability due to sine-Gaussians. This is found in exactly
the same way as the sensitivity to ring-downs but using
sine-Gaussian injections. The false alarm probability due to
sine-Gaussians, therefore, is simply the fraction of the 250 injected sine-Gaussians
which cause the odds ratio to exceed the threshold.

\subsection{$\textbf{O}_{\textrm{\scriptsize{12}}}$:  ring-down vs noise}
\begin{center}
\begin{figure}[]
    \includegraphics{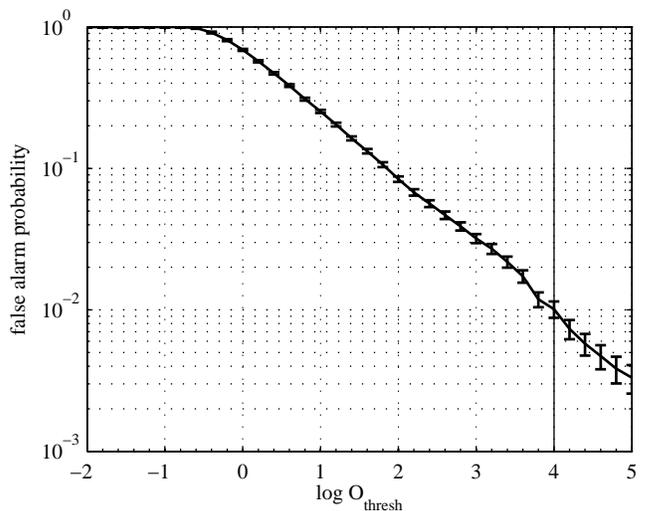}
    \caption[False alarm probabilities in white noise for $O_{12}$]{The false alarm
probabilities
    in Gaussian white noise for different $\log_{10} O_{12}$ thresholds.  Error bars
represent $1\sigma$ Poissonian standard errors.  The solid vertical
line marks the threshold applied when using $\log_{10} O_{12}$.}
    \label{fig:noise_thresh}
\end{figure}
\end{center}

We take our desired false alarm probability here to be $1\%$.
Fig.~\ref{fig:noise_thresh} shows the false alarm probability as a
function of the log odds ratio threshold, $\log_{10} \Othresh$, and we
see that a false alarm probability of $1\%$ is given by $\log_{10}
\Othresh = 4$.

\begin{center}
\begin{figure}[]
    \includegraphics{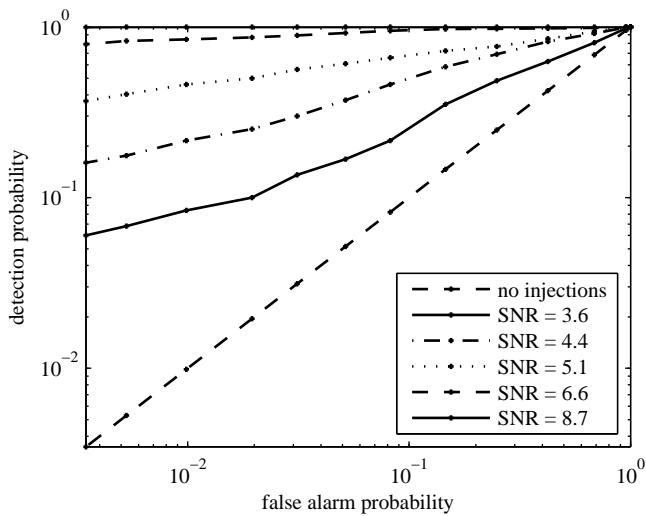}
    \caption[ROC curve for $O_{12}$]{Receiver operating characteristic curves for
$O_{12}$.  False alarm probabilities are those from amplitude fluctuations in
Gaussian white noise.}
    \label{fig:ROC_O12}
\end{figure}
\end{center}

Fig.~\ref{fig:ROC_O12} shows \emph{receiver operating
characteristic} (ROC) curves for $O_{12}$. ROC curves are plots of
sensitivity (the probability of detecting what we are looking for)
as a function of false alarm probability (the probability of
claiming a detection due to a spurious noise event or instrumental
glitch) for a given strength signal.  The different false alarm
probabilities correspond to different values of the detection
threshold, $\Othresh$, and are found by reading the appropriate
values from Fig.~\ref{fig:noise_thresh}.  When there are no injected
signals, the only events to cause the odds ratio to cross the
threshold are false alarms and the false alarm probability should be
equal to the detection probability.

\begin{center}
\begin{figure}[]
    \includegraphics{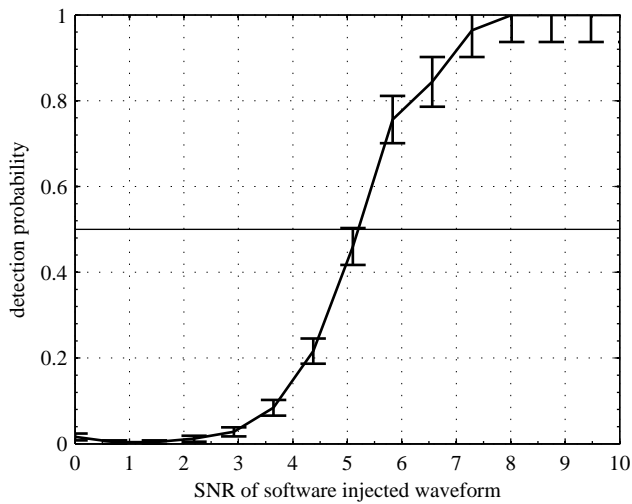}
    \caption[Efficiency curve for $O_{12}$]{The detection efficiency obtained for
$\log_{10} O_{12}$ using a threshold $\log_{10} O_{\textrm {\scriptsize thresh}} = 4$.
Error bars show $1\sigma$ standard Possonian errors.}
    \label{fig:O12_efficiency}
\end{figure}
\end{center}

Although the ROC curves give some indication of how the sensitivity
varies with injected signal strength, it is also helpful to examine
efficiency curves, where the fraction of detected ring-downs is
plotted against the injected signal strength, for a given threshold.
Fig.~\ref{fig:O12_efficiency} shows the detection efficiency using
$\log_{10} \Othresh = 4$ with corresponding false alarm probability
of $1\%$.

The algorithm's performance is best summarised by the
signal-to-noise ratio required to give a detection probability of
$50\%$ while maintaining the desired false alarm probability. A
signal-to-noise ratio $\rho = 5.2$ is required for $50\%$ ring-down
detection, corresponding to an initial ring-down amplitude at
Earth of $h_0 = 1.8 \times 10^{-21}$ at current LIGO sensitivities.

Equation \ref{eqn:GWenergy} on page \pageref{eqn:GWenergy} relates $h_0$ to the distance
to the source and the energy emitted as gravitational waves. Using this equation and
assuming the fiducial distance of $15$ kpc, the energy required to generate this initial
ring-down amplitude at the Earth is $1.3\times10^{-5}M_{\astrosun}c^2$. 
Conversely, when we assume the fiducial energy in \gws of $10^{-11}M_{\astrosun}$, the
distance to the source must be $13.2$ pc for this amplitude.  These results are summarised
in Table \ref{table:results}.

\subsection{$\textbf{O}_{\textrm{\scriptsize{123}}}$:  Ring-down vs noise or
sine-Gaussian glitch}
 We now evaluate the performance of the algorithm using the more
robust comparison between ring-downs, white noise and
sine-Gaussians.

The response of $O_{123}$ to sine-Gaussians is compared with the
response to ring-down injections in Fig.~\ref{fig:meanO123}.  The
horizontal axis shows the signal-to-noise ratio of each type of
injected signal (ring-down or sine-Gaussian) and the vertical axis
gives the mean value of $\log_{10} O_{123}$, calculated from 250
injections of each signal type.  The response can be separated into
three regions of signal-to-noise ratio, $\rho$:
\begin{center}
\begin{figure}[]
    \includegraphics{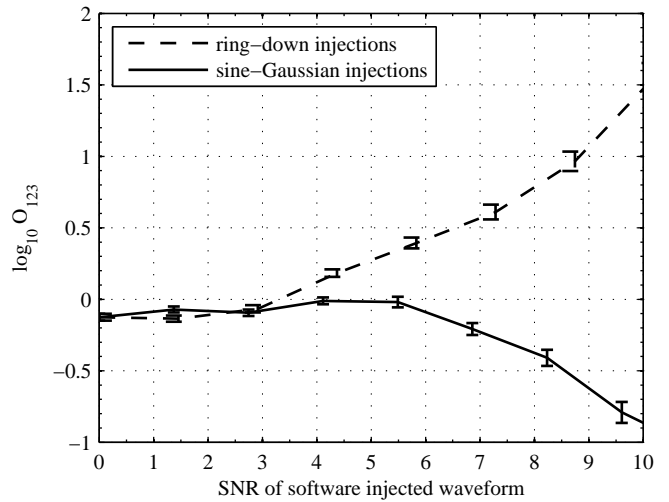}
    \caption[Mean $O_{123}$ for ring-downs and sine-Gaussians]{The mean value of $\log_{10}
O_{123}$ for
    varying strengths of signal-to-noise ratio.  Results of ring-down injections are shown
by the
    dashed curve, while the solid curve shows the results of the sine-Gaussian
injections.  Error bars represent $1\sigma$ standard errors in the mean values of $\log_{10} O_{123}$.}
    \label{fig:meanO123}
\end{figure}
\end{center}
\begin{description}
\item[{$\rho \lesssim 3$}]: there is no response to
sine-Gaussians or
ring-downs and any signal present is indistinguishable from noise.

\item[{$3 \lesssim \rho \lesssim 5$}]: the algorithm begins to respond to the
 ring-down injections.  The sine-Gaussian injections generate a very slight rise in the log odds.

\item[{$\rho \gtrsim 5$}]: the odds in favour of a ring-down
begin to grow for the ring-down injections and rapidly falls off for the sine-Gaussian injections.
\end{description}

\begin{center}
\begin{figure}[]
    \includegraphics{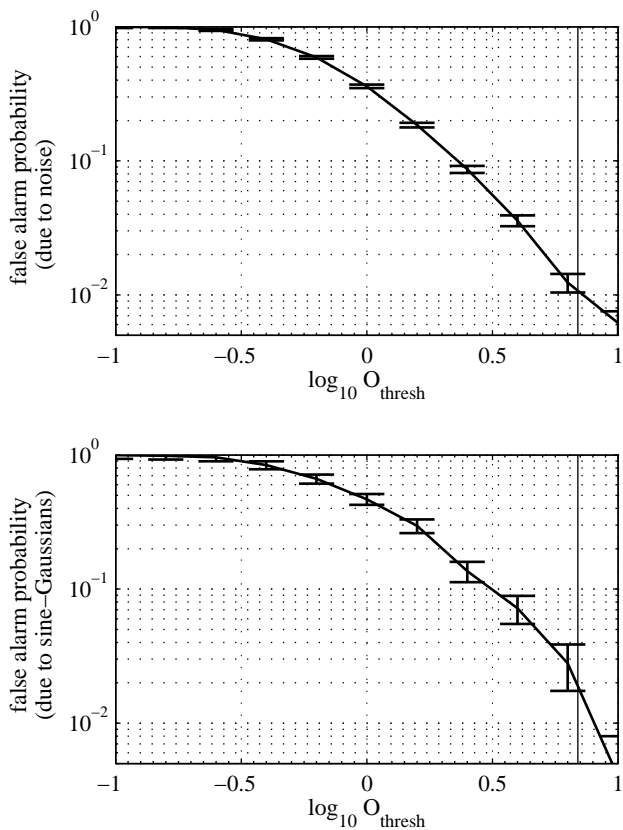}
    \caption[Sine-Gaussian false alarm probabilities for $O_{123}$]{
\emph{Top panel}: false alarm probabilities and corresponding $\log_{10} O_{123}$ thresholds
for Gaussian white noise.  \emph{Bottom panel}: false alarm probabilities and corresponding $\log_{10} O_{123}$ thresholds
for sine-Gaussians with signal-to-noise ratio $\rho \sim 4$.  In each case, the solid vertical line indicates the threshold required for a $1\%$
 false alarm probability in Gaussian white noise.  Error bars show $1\sigma$ Poissonian standard errors.}
    \label{fig:SG_thresh}
\end{figure}
\end{center}

We use the intermediate region (i.e., where $3 \lesssim \rho \lesssim 5$) to assume a worst-case
scenario where all sine-Gaussian glitches have a signal-to-noise ratio $\rho \sim 4$ and compute
the false alarm probability from sine-Gaussians for different odds ratio thresholds.  The results 
are shown in Fig.~\ref{fig:SG_thresh}.

We now find that the threshold required to give a false alarm probability of $1\%$ in
Gaussian white noise is $\log_{10} \Othresh=0.84$. This corresponds to a false alarm 
probability of $10\%$ when the data contains a sine-Gaussian. We note two points here:

\begin{itemize}
\item [a)] we have assumed the presence of sine-Gaussian glitches \emph{a priori}.  While it is appropriate
to assume the presence of Gaussian white noise, we do not have a population
model for the sine-Gaussian glitches. A more informative estimate
of the probability of mistakenly detecting a sine-Gaussian glitch
should fold in the effects of such a population model through the prior on $M_3$. 
In this work, we are more interested in demonstrating the inherent ability of the
algorithm to discriminate between similar waveforms and this will not be considered.

\item [b)] Only sine-Gaussians with signal-to-noise ratios of $\rho \sim 4$
are considered. Away from this value,
the odds ratio drops off rapidly so that the $10\%$
false alarm probability can be regarded as an upper
limit for sine-Gaussian glitches.
\end{itemize}

\begin{center}
\begin{figure}[]
    \includegraphics{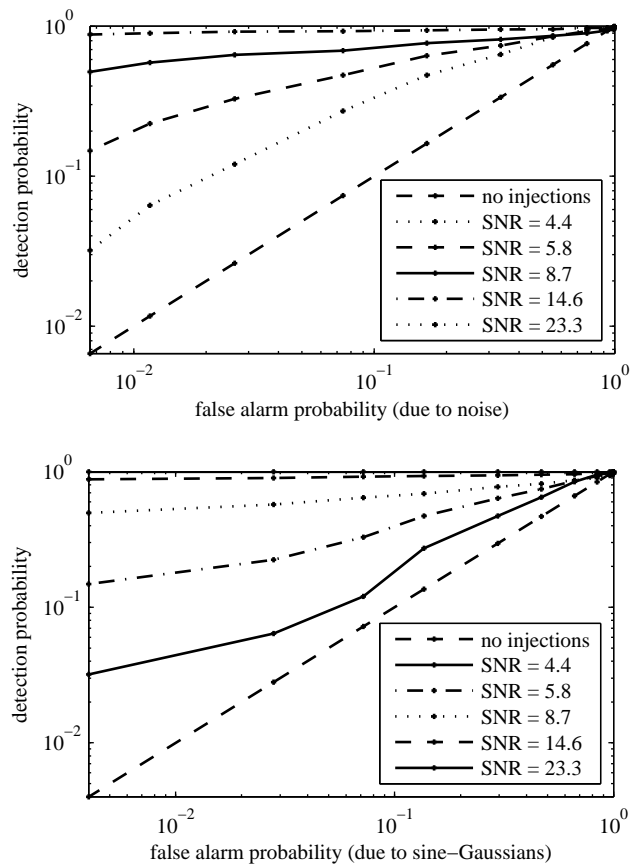}
    \caption[ROC curve 1]{Receiver operating characteristic curves for $O_{123}$.
\emph{Top panel}: ROC curve with
    false alarm probabilities due to spurious noise events.  \emph{Bottom panel}: ROC
curve with
    false alarm probabilities due to sine-Gaussian glitches.}
    \label{fig:ROC}
\end{figure}
\end{center}

Because we have two possible sources of false alarms, sine-Gaussians
and noise events, we produce ROC figures for each, shown in each of
the panels of Fig.~\ref{fig:ROC}.

\begin{center}
\begin{figure}[]
    \includegraphics{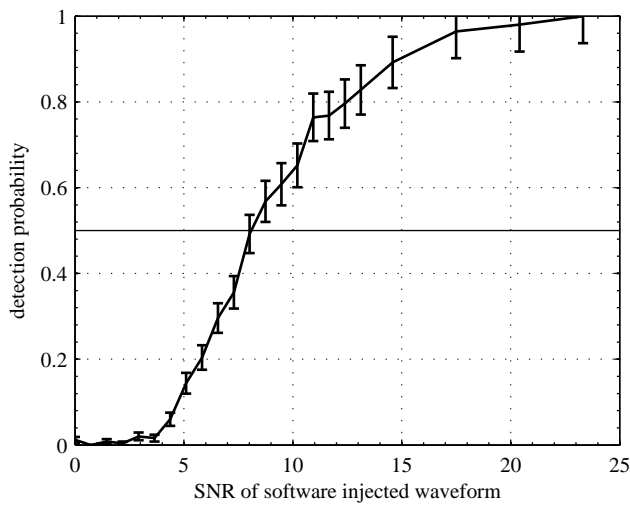}
    \caption[Efficiency curve for $O_{123}$]{The detection efficiency obtained for
$\log_{10} O_{123}$ using a threshold $\log_{10} O_{\textrm {\scriptsize thresh}} = 0.84$.
The solid horizontal line indicates $50\%$ detection efficiency.
Error bars show $1\sigma$ Poissonian standard errors.}
    \label{fig:O123_efficiency}
\end{figure}
\end{center}

For comparison with the sensitivity of $O_{12}$
Fig.~\ref{fig:O123_efficiency} shows the detection efficiency for
$O_{123}$.  We find that the signal-to-noise ratio required for $50\%$ detection
efficiency and $1\%$ false alarm probability in Gaussian white noise
is $\rho^{50\%}=8.0$.

Again, it is important to consider these results in an astrophysical
context.  With the noise amplitude spectral density used for these
investigations ($10^{-22}$ Hz$^{-1/2}$), a ring-down signal-to-noise
ratio $\rho = 8.0$ corresponds to an initial amplitude $h_0 = 2.9
\times 10^{-21}$ Hz$^{-1/2}$. The gravitational wave energy required
to produce a ring-down signal with $50\%$ detection probability at
the Earth with this amplitude is $ 3.3 \times
10^{-5}M_{\astrosun}c^2$ for a source at $15$ kpc.  Similarly, if we assume that
$10^{-11}M_{\astrosun}c^2$ is emitted as \gws, the source must lie at $8.2$ pc. These
results are summarised and compared with those from $O_{12}$ in Table
\ref{table:results}.

\begin{center}
\begin{table*}
\begin{ruledtabular}
\begin{tabular}{l c c c}
Parameter & Symbol & Value Using $O_{12}$ & Value Using $O_{123}$\\ \hline
Log odds threshold & $\log_{10} \Othresh$ & $4$ &$0.84$ \\
False alarm probability (due to noise) & $p(-|N)$ & $1\%$ &$1\%$ \\
False alarm probability (due to sine-Gaussians) & $p(-|SG)$ & $-$ &$10 \%$\\
Signal-to-noise & $\rho^{50\%}$ & $5.2$ &$8.0$\\
Initial strain amplitude & $h_0^{50\%}$ & $1.8\times10^{-21}$ &$2.9\times 10^{-21}$ \\
GW Energy & $E\subgw^{50\%}$ & $1.3\times10^{-5}M_{\astrosun}c^2$ &$3.3\times
10^{-5}M_{\astrosun}c^2$ \\
Range & $D\subgw^{50\%}$ & $13.2$ pc & $8.2$ pc \\
Root-sum-squared strain & $h_{rss}^{50\%}$ &$4.1\times10^{-22}$ & $6.6\times 10^{-22}$\\
\hline
\end{tabular}
\end{ruledtabular}
\caption[Results]{\label{table:results}Simulated sensitivity
estimates for $50\%$ detection efficiency and corresponding false
alarm probabilities.  $p(-|N)$ is the probability of a false alarm
from white noise for the given threshold $\log_{10} \Othresh$.
Similarly, $p(-|SG)$ is the probability of a false alarm given a
sine-Gaussian event.  All estimates assume a ring-down frequency
$\nu = 2$\,kHz, decay timescale $\tau = 207.5$ ms and a noise
amplitude spectral density $10^{-22}$ Hz$^{-1/2}$. $E\subgw^{50\%}$
is calculated for a distance $D=15$\,kpc and $D\subgw^{50\%}$ is calculated for an energy
in \gws of $10^{-11}M_{\astrosun}c^2$.}
\end{table*}
\end{center}

\subsection{Comparison to matched filtering}\label{sec:matched-filtering}

 The LIGO Algorithm Library (LAL) \cite{LAL}  and LALapps 
software repositories used for much of the gravitational
wave data analysis within the LIGO Scientific Collabora-
tion (LSC)\footnote{\url{http://www.ligo.org}} contain software for performing a matched
filter based search for ring-down signals. Currently this is
being used to search for ring-downs from perturbed black
holes \cite{Goggin:2006}. The software can be used to compare the false
alarm probability of the matched filtering method with
our method by running it on simulated white noise for
the range of parameters given in  Table~\ref{table:priors}
\footnote{The matched filtering code defines the
ring-down in terms of frequency and quality factor, $Q$, 
where $Q = \pi \tau f$.}. The template
bank for the matched filtering was produced to give a
maximum mismatch between adjacent templates of $1\%$.

We define the false alarm probability to be the fraction
of detection candidates generated by the algorithm whose
signal-to-noise ratio crosses a given threshold. Fig. 12
shows the false alarm probability as a function of signal-
to-noise ratio threshold: for a false alarm probabilty of
$1\%$, we require a signal-to-noise ratio threshold of $5.85$ (compared with $\rho = 5.2$
for the evidence based search).

It may seem surprising that the evidence-based search marginally
out-performs matched filtering.  It is important to remember, however, that
the approaches are by no means equivalent:
matched filtering searches for consistency with a given waveform, whereas the evidence-based
approach searches for both consistency with the same waveform but also for \emph{inconsistency}
with a noise model.  Since the noise model is highly sensitive to excess power (which exponentially
decreases its likelihood), it does not seem unreasonable that there is a slight disparity
between the approaches.

\begin{center}
\begin{figure}[]
    \includegraphics{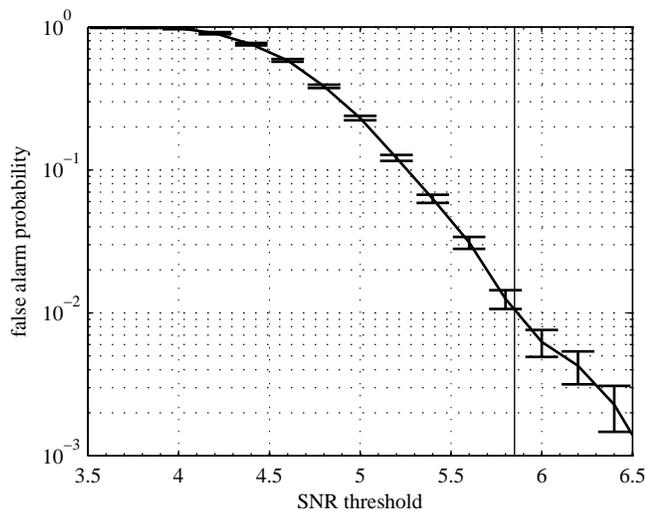}
    \caption[Matched filter triggers]{The cumulative percentage of triggers produced by a 
matched filter based ring-down signal search code over a range of signal-to-noise ratios
for data
consisting of simulated Gaussian white noise.  Error bars indicate 1$\sigma$ Poissonian
standard errors.  The solid vertical line indicates the $\rho = 5.85$ threshold for a $1\%$
false alarm probability.}
    \label{fig:matchedfilter}
\end{figure}
\end{center}
%

\section{Further extensions}
We now highlight some outstanding issues and possible important extensions to
this work which may be required to make the transition from
`proof-of-concept' to a useful search tool. Particularly, we address
the issues of constructing the catalogue of alternative hypotheses,
the extension to a multi-detector analysis and the outline for a
full analysis pipeline.

\subsection*{Glitch classification}
It is likely that the inclusion of extra information regarding
glitches will increase robustness against instrumental transients
which would otherwise generate a candidate detection event.  For
this to be effective, we require some classification scheme for
these instrumental glitches which would allow the construction of a
catalogue of glitch models as well as their associated priors.

Essentially, what is required is an automated pattern recognition
tool which could be `trained' on off-source interferometer data and
used to generate a generic library of transient features.
Fortunately, on-going detector characterisation work aims to perform 
such an analysis \cite{soma:glitches}.

\subsection*{Multiple detector case}
A potential issue with this search is in fact the prevalence of
instrumental ring-downs already present in the interferometer data.
Unless such features can be vetoed using known instrumental
couplings, for example \cite{ajith:2006}, the only way to
distinguish these from the targeted gravitational wave ring-down is
to search for coincidences between multiple detectors.

Here, we benefit again from the simplicity of the Bayesian
formalism.  In the single detector case outlined in this work, we
aim to compute the posterior probability for some gravitational wave
model, given the interferometer data and some background
information.  In the multi-detector case we still seek the posterior
probability for some model but now using the information contained
in each detector's output.  Suppose then that we have some
gravitational wave model $M\subgw$ and detector outputs $D_1$ and
$D_2$.  We can immediately write down the posterior for $M\subgw$
using Bayes' theorem
\begin{equation}\label{eq:multiBayes}
    p(M\subgw|D_1,D_2,I) = \frac{p(M\subgw|I)p(D_1,D_2|M\subgw,I)}{p(D_1,D_2|I)},
\end{equation}
where the gravitational wave model $M\subgw$ factors in the
appropriate detector response functions for the source sky position
and for detectors with uncorrelated output, the joint probabilities
are the products of the individual probabilities. For $N$ detectors
\begin{equation}\label{eq:multiBayes_joint}
    p(M\subgw|\{D\},I) = p(M\subgw|I)\prod_{i=1}^N \frac{p(D_i|M\subgw,I)}{p(D_i|I)}
\end{equation}
and we can again eliminate the denominator by comparing the relative
probabilities of different models via the odds ratio.  Notice,
however, that if the probability from one detector is very large but
low in the other detectors it is still possible to get a high
posterior across all the detectors.  It would, therefore, be
sensible to apply a cut to the data such that the odds must cross
some threshold for each individual detector before considering the
multi-detector case.
\subsection*{Analysis pipeline}
\begin{center}
\begin{figure}[]
    \scalebox{0.4}{\includegraphics{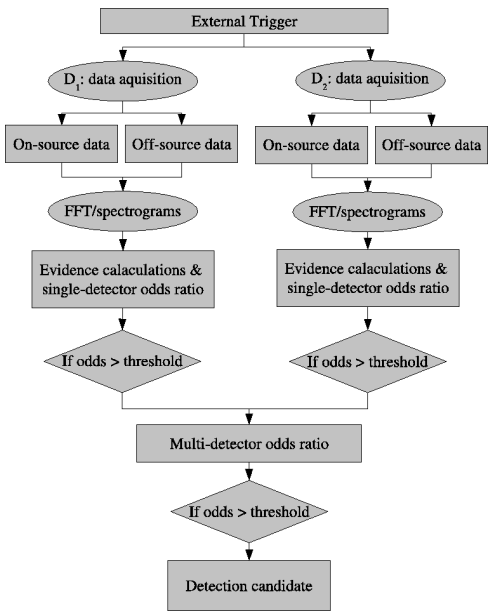}}
    \caption[Analysis Pipeline]{Planned multi-detector, evidence-based analysis pipeline.}
    \label{fig:pipeline}
\end{figure}
\end{center}

Finally, we present the outline for a planned analysis pipeline in
Fig.~\ref{fig:pipeline}.  Here, we assume a network of two detectors
and that there is no correlation in the output from each.
Upon reception of an external trigger, such as a pulsar glitch or
soft $\gamma$-ray repeater flare, we would retrieve interferometer
data from near the event (on-source data).  The length of data
required will typically depend on the time resolution of the
external trigger. Additionally, we require data preceding the
external trigger to provide some estimate of the background odds
ratio and hence set the threshold for the desired false alarm
probability.

Following the acquisition of both on and off-source data for each
detector, we transform to the Fourier domain and construct
spectrograms of each.   From here, the methods outlined in this work
can be used to compute the evidences for each gravitational wave and
glitch model and the odds ratio can be computed for each detector
using on and off-source data.

If the odds in \emph{both} detectors cross some appropriate
threshold, then we can go on to calculate the multi-detector odds
ratio.  If this multi-detector odds ratio also crosses a threshold
set from off-source data then we have a candidate detection.

Alternatively, if the odds ratio from just one of the detectors in
the network crosses the respective threshold, then there is probably
good evidence for a glitch. Further examination of the odds ratios
for each glitch model may reveal its nature and the information
could conceivably be used to update the prior on that particular
model in future studies.


\section{Concluding remarks}
We have examined the feasibility of using a Bayesian evidence
discriminator in an externally-triggered search for gravitational
waves produced by neutron star quasi-normal mode ring-downs.  The
evidence may be thought of as the total probability of the data
given some model. By comparing the evidences for competing models
via the `odds ratio', we can naturally make maximal use of prior
information regarding the source as well as any information
available from detector characterisation studies.

This is particularly important in the single-detector case.  Here
there may be instrumental transients or `glitches' which would
otherwise generate a detection candidate in algorithms which only
look for consistency with the target model.  We begin with a simple
formulation of the odds ratio, $O_{12}$, where we compare the
probability of a noisy ring-down model with the probability of noise
alone.  We find this yields a $50\%$ detection efficiency for
ring-downs with frequency $2$\,kHz, decay time $0.21$\,s and signal-to-noise ratio
$\rho=5.2$ for a $1\%$ false alarm
probability in Gaussian white noise of amplitude spectral density
$10^{-22}$ Hz$^{-1/2}$.  To obtain the same false alarm probability using matched
filtering we require a signal-to-noise ratio threshold $\rho=5.85$.  As described
in section \ref{sec:matched-filtering} this disparity is most likely due to the
fundamental difference between using the odds ratio, which compares the evidences from
two or more competing models and matched filtering, which only searches for consistency
with a single target waveform.

Assuming $10^{-11}M_{\astrosun}c^2$ is emitted as gravitational waves, 
these sources are observable to a distance of $13.2$\,pc.  Conversely, a
source at a distance of $15$\,kpc requires an energy of $1.3\times10^{-5}M_{\astrosun}
c^2$ to be emitted as gravitational waves. This is several orders of magnitude
greater than the typical energy we might expect from a pulsar
glitch, making these sources a more attractive target for
next-generation detectors such as advanced LIGO \cite{Fritschel:2003} and GEO-HF
\cite{GEOHF}.  If, for example, we consider advanced LIGO
with a factor $\sim 10$ improvement in sensitivity over LIGO, the energy required to
produce the same signal-to-noise ratio drops to $\sim 1.0\times10^{-7}M_{\astrosun}c^2$.
While this is still well below our fiducial pulsar glitch \gw energy of
$10^{-11}M_{\astrosun}c^2$, it is comparable to the \gw energy expected to be emitted
following the axisymmetric collapse of the core of a massive star, some fraction of which
will be channelled into the oscillatory modes of the proto-neutron star
\cite{SNe_ringdowns, dimmelmeier}.  Again, if we assume the fiducial \gw energy of
$10^{-11}M_{\astrosun}c^2$, advanced LIGO will be sensitive to neutron star ring-downs at
a distance of $148$\,pc.

We find the simple comparison between white noise and ring-downs is
insufficient to discriminate between different types of signal and
will be fooled by any transient departure from Gaussianity in the
noise. Robustness is improved by including a toy catalogue of glitch
models (a sine-Gaussian). The odds ratio may then be extended (i.e.,
$O_{123}$) to mitigate the effects of these unwanted signals at a
relatively small sensitivity cost: a ring-down signal-to-noise ratio
of $\rho=8.0$ is now required for the $50\%$ ring-down detection
efficiency and $1\%$ false alarm probability due to Gaussian white
noise, so that the total gravitational wave energy required for
$50\%$ detection probability of a source at $15$\,kpc is now
$3.3\times 10^{-5}M_{\astrosun} c^2$ and the observable range is now  $8.2$\,pc.  With
advanced LIGO sensitivity, these estimates improve to $2.7\times10^{-8}M_{\astrosun}c^2$
and $91.6$\,pc for the energy in \gws and the observable range, respectively.  We also
 find that this extened odds ratio only has a $10\%$ chance of falsely identifying a
 sine-Gaussian as a ring-down.  This assumes a worst-case scenario in which the
 sine-Gaussian has a signal-to-noise ratio $\rho\sim4$ (where the algorithm's response
 to these signals peaks).  This is therefore an upper limit so that the 
 false alarm rate from sine-Gaussians will generally be much lower.

Finally, we have made mention of some of the future work required to
use the methodology in this work in a search for gravitational
waves.  For the single detector case, it will be necessary to
acquire an appropriate catalogue of glitch models and associated
prior probabilities for their parameters.  The extension to the
multi-detector case will require considerable modification to the
waveform model to account for light travel time between detectors
and the appropriate antenna response functions for a given sky
location. For this search, we are at an advantage in that we know
the source location and event time, and hence the factors introduced
by the antenna response and time delay are known.  We note, however,
that if this was not so, as would be the case for an all-sky search
or where we had a trigger but no source location, it would be
necessary to marginalise over the source location which would
significantly complicate the evidence calculation.

With a clear idea of the future analysis pipeline, the next step is
to characterise the algorithm response using real interferometer
data and all the consequences of non-Gaussian noise.


\begin{acknowledgments}
The authors are very grateful to the LIGO scientific collaboration
for their support.  We are indebted to many of our colleagues for
their fruitful discussion and advice.  In particular, we would like
to thank Patrick Sutton and Szabolcs Marka for their valuable
comments on the manuscript and John Veitch for his insight and
lively discussion regarding the subject matter.  This work has been
supported by the UK Particle Physics and Astronomy Research Council.
\end{acknowledgments}
\bibliography{references}

\begin{thebibliography}{28}
\expandafter\ifx\csname natexlab\endcsname\relax\def\natexlab#1{#1}\fi
\expandafter\ifx\csname bibnamefont\endcsname\relax
  \def\bibnamefont#1{#1}\fi
\expandafter\ifx\csname bibfnamefont\endcsname\relax
  \def\bibfnamefont#1{#1}\fi
\expandafter\ifx\csname citenamefont\endcsname\relax
  \def\citenamefont#1{#1}\fi
\expandafter\ifx\csname url\endcsname\relax
  \def\url#1{\texttt{#1}}\fi
\expandafter\ifx\csname urlprefix\endcsname\relax\def\urlprefix{URL }\fi
\providecommand{\bibinfo}[2]{#2}
\providecommand{\eprint}[2][]{\url{#2}}

\bibitem[{\citenamefont{{Thorne}}(1969)}]{thorne:1969}
\bibinfo{author}{\bibfnamefont{K.~S.} \bibnamefont{{Thorne}}},
  \bibinfo{journal}{Astrophys. J.} \textbf{\bibinfo{volume}{158}},
  \bibinfo{pages}{1} (\bibinfo{year}{1969}).

\bibitem[{\citenamefont{{Kokkotas} et~al.}(2001)\citenamefont{{Kokkotas},
  {Apostolatos}, and {Andersson}}}]{andersson_fingerprints}
\bibinfo{author}{\bibfnamefont{K.~D.} \bibnamefont{{Kokkotas}}},
  \bibinfo{author}{\bibfnamefont{T.~A.} \bibnamefont{{Apostolatos}}},
  \bibnamefont{and}
  \bibinfo{author}{\bibfnamefont{N.}~\bibnamefont{{Andersson}}},
  \bibinfo{journal}{Mon. Not. R. Astron. Soc.} \textbf{\bibinfo{volume}{320}},
  \bibinfo{pages}{307} (\bibinfo{year}{2001}), \eprint{gr-qc/9901072}.

\bibitem[{\citenamefont{{de Freitas Pacheco}}(1998)}]{pacheco}
\bibinfo{author}{\bibfnamefont{J.~A.} \bibnamefont{{de Freitas Pacheco}}},
  \bibinfo{journal}{Astron. Astrophys.} \textbf{\bibinfo{volume}{336}},
  \bibinfo{pages}{397} (\bibinfo{year}{1998}),
  \eprint{\href{http://www.arxiv.org/abs/astro-ph/9805321}{astro-ph/9805321}}.

\bibitem[{\citenamefont{{Andersson} and {Kokkotas}}(1998)}]{andersson}
\bibinfo{author}{\bibfnamefont{N.}~\bibnamefont{{Andersson}}} \bibnamefont{and}
  \bibinfo{author}{\bibfnamefont{K.}~\bibnamefont{{Kokkotas}}},
  \bibinfo{journal}{Mon. Not. R. Astron. Soc.} \textbf{\bibinfo{volume}{299}},
  \bibinfo{pages}{1059} (\bibinfo{year}{1998}),
  \eprint{\href{http://www.arxiv.org/abs/gr-qc/9711088}{gr-qc/9711088}}.

\bibitem[{\citenamefont{Andersson}(2003)}]{andersson_review}
\bibinfo{author}{\bibfnamefont{N.}~\bibnamefont{Andersson}},
  \bibinfo{journal}{Classical and Quantum Gravity}
  \textbf{\bibinfo{volume}{20}}, \bibinfo{pages}{R105} (\bibinfo{year}{2003}),
  \urlprefix\url{http://stacks.iop.org/0264-9381/20/R105}.

\bibitem[{\citenamefont{{L{\"u}ck} et~al.}(2006)\citenamefont{{L{\"u}ck},
  {Hewitson}, {Ajith}, {Allen}, {Aufmuth}, {Aulbert}, {Babak},
  {Balasubramanian}, {Barr}, {Berukoff} et~al.}}]{GEO}
\bibinfo{author}{\bibfnamefont{H.}~\bibnamefont{{L{\"u}ck}}},
  \bibinfo{author}{\bibfnamefont{M.}~\bibnamefont{{Hewitson}}},
  \bibinfo{author}{\bibfnamefont{P.}~\bibnamefont{{Ajith}}},
  \bibinfo{author}{\bibfnamefont{B.}~\bibnamefont{{Allen}}},
  \bibinfo{author}{\bibfnamefont{P.}~\bibnamefont{{Aufmuth}}},
  \bibinfo{author}{\bibfnamefont{C.}~\bibnamefont{{Aulbert}}},
  \bibinfo{author}{\bibfnamefont{S.}~\bibnamefont{{Babak}}},
  \bibinfo{author}{\bibfnamefont{R.}~\bibnamefont{{Balasubramanian}}},
  \bibinfo{author}{\bibfnamefont{B.~W.} \bibnamefont{{Barr}}},
  \bibinfo{author}{\bibfnamefont{S.}~\bibnamefont{{Berukoff}}},
  \bibnamefont{et~al.}, \bibinfo{journal}{Classical and Quantum Gravity}
  \textbf{\bibinfo{volume}{23}}, \bibinfo{pages}{71} (\bibinfo{year}{2006}).

\bibitem[{\citenamefont{{Sigg (for the LIGO Science
  Collaboration)}}(2006)}]{LIGO}
\bibinfo{author}{\bibfnamefont{D.}~\bibnamefont{{Sigg (for the LIGO Science
  Collaboration)}}}, \bibinfo{journal}{Classical and Quantum Gravity}
  \textbf{\bibinfo{volume}{23}}, \bibinfo{pages}{S51} (\bibinfo{year}{2006}),
  \urlprefix\url{http://stacks.iop.org/0264-9381/23/S51}.

\bibitem[{\citenamefont{{Tournefier} and {VIRGO Collaboration}}(2005)}]{VIRGO}
\bibinfo{author}{\bibfnamefont{E.}~\bibnamefont{{Tournefier}}}
  \bibnamefont{and} \bibinfo{author}{\bibnamefont{{VIRGO Collaboration}}}, in
  \emph{\bibinfo{booktitle}{SF2A-2005: Semaine de l'Astrophysique Francaise}},
  edited by \bibinfo{editor}{\bibfnamefont{F.}~\bibnamefont{{Casoli}}},
  \bibinfo{editor}{\bibfnamefont{T.}~\bibnamefont{{Contini}}},
  \bibinfo{editor}{\bibfnamefont{J.~M.} \bibnamefont{{Hameury}}},
  \bibnamefont{and} \bibinfo{editor}{\bibfnamefont{L.}~\bibnamefont{{Pagani}}}
  (\bibinfo{year}{2005}), p. \bibinfo{pages}{539}.

\bibitem[{\citenamefont{{Samuelsson} and {Andersson}}(2007)}]{Samuelsson:2006}
\bibinfo{author}{\bibfnamefont{L.}~\bibnamefont{{Samuelsson}}}
  \bibnamefont{and}
  \bibinfo{author}{\bibfnamefont{N.}~\bibnamefont{{Andersson}}},
  \bibinfo{journal}{Mon. Not. R. Astron. Soc.} \textbf{\bibinfo{volume}{374}},
  \bibinfo{pages}{256} (\bibinfo{year}{2007}),
  \eprint{\href{http://www.arxiv.org/abs/astro-ph/0609265}{astro-ph/0609265}}.

\bibitem[{\citenamefont{{Benhar} et~al.}(2004)\citenamefont{{Benhar},
  {Ferrari}, and {Gualtieri}}}]{benhar}
\bibinfo{author}{\bibfnamefont{O.}~\bibnamefont{{Benhar}}},
  \bibinfo{author}{\bibfnamefont{V.}~\bibnamefont{{Ferrari}}},
  \bibnamefont{and}
  \bibinfo{author}{\bibfnamefont{L.}~\bibnamefont{{Gualtieri}}},
  \bibinfo{journal}{Phys. Rev. D} \textbf{\bibinfo{volume}{70}},
  \bibinfo{pages}{124015} (\bibinfo{year}{2004}),
  \eprint{\href{http://www.arxiv.org/abs/astro-ph/0407529}{astro-ph/0407529}}.

\bibitem[{\citenamefont{{Middleditch} et~al.}(2006)\citenamefont{{Middleditch},
  {Marshall}, {Wang}, {Gotthelf}, and {Zhang}}}]{Middleditch:2006}
\bibinfo{author}{\bibfnamefont{J.}~\bibnamefont{{Middleditch}}},
  \bibinfo{author}{\bibfnamefont{F.~E.} \bibnamefont{{Marshall}}},
  \bibinfo{author}{\bibfnamefont{Q.~D.} \bibnamefont{{Wang}}},
  \bibinfo{author}{\bibfnamefont{E.~V.} \bibnamefont{{Gotthelf}}},
  \bibnamefont{and} \bibinfo{author}{\bibfnamefont{W.}~\bibnamefont{{Zhang}}},
  \bibinfo{journal}{\apj} \textbf{\bibinfo{volume}{652}}, \bibinfo{pages}{1531}
  (\bibinfo{year}{2006}), \eprint{astro-ph/0605007}.

\bibitem[{Cra()}]{CrabEphemeris}
\emph{\bibinfo{title}{{Jodrell Bank Crab Pulsar Monthly Ephemeris {\tt
  \url{http://www.jb.man.ac.uk/~pulsar/crab.html}}}}}.

\bibitem[{\citenamefont{{Dodson} et~al.}(2006)\citenamefont{{Dodson}, {Lewis},
  and {McCulloch}}}]{Dodson:2006}
\bibinfo{author}{\bibfnamefont{R.}~\bibnamefont{{Dodson}}},
  \bibinfo{author}{\bibfnamefont{D.}~\bibnamefont{{Lewis}}}, \bibnamefont{and}
  \bibinfo{author}{\bibfnamefont{P.}~\bibnamefont{{McCulloch}}},
  \bibinfo{journal}{ArXiv Astrophysics e-prints}  (\bibinfo{year}{2006}),
  \eprint{astro-ph/0612371}.

\bibitem[{\citenamefont{{Nakagawa} et~al.}(2007)\citenamefont{{Nakagawa},
  {Yoshida}, {Hurley}, {Atteia}, {Maetou}, {Tamagawa}, {Suzuki}, {Yamazaki},
  {Tanaka}, {Kawai} et~al.}}]{Nakagawa:2007}
\bibinfo{author}{\bibfnamefont{Y.~E.} \bibnamefont{{Nakagawa}}},
  \bibinfo{author}{\bibfnamefont{A.}~\bibnamefont{{Yoshida}}},
  \bibinfo{author}{\bibfnamefont{K.}~\bibnamefont{{Hurley}}},
  \bibinfo{author}{\bibfnamefont{J.-L.} \bibnamefont{{Atteia}}},
  \bibinfo{author}{\bibfnamefont{M.}~\bibnamefont{{Maetou}}},
  \bibinfo{author}{\bibfnamefont{T.}~\bibnamefont{{Tamagawa}}},
  \bibinfo{author}{\bibfnamefont{M.}~\bibnamefont{{Suzuki}}},
  \bibinfo{author}{\bibfnamefont{T.}~\bibnamefont{{Yamazaki}}},
  \bibinfo{author}{\bibfnamefont{K.}~\bibnamefont{{Tanaka}}},
  \bibinfo{author}{\bibfnamefont{N.}~\bibnamefont{{Kawai}}},
  \bibnamefont{et~al.}, \bibinfo{journal}{ArXiv Astrophysics e-prints}
  (\bibinfo{year}{2007}), \eprint{astro-ph/0701701}.

\bibitem[{\citenamefont{{Cheng} et~al.}(1988)\citenamefont{{Cheng}, {Pines},
  {Alpar}, and {Shaham}}}]{Cheng:1988}
\bibinfo{author}{\bibfnamefont{K.~S.} \bibnamefont{{Cheng}}},
  \bibinfo{author}{\bibfnamefont{D.}~\bibnamefont{{Pines}}},
  \bibinfo{author}{\bibfnamefont{M.~A.} \bibnamefont{{Alpar}}},
  \bibnamefont{and} \bibinfo{author}{\bibfnamefont{J.}~\bibnamefont{{Shaham}}},
  \bibinfo{journal}{\apj} \textbf{\bibinfo{volume}{330}}, \bibinfo{pages}{835}
  (\bibinfo{year}{1988}).

\bibitem[{\citenamefont{{Franco} et~al.}(2000)\citenamefont{{Franco}, {Link},
  and {Epstein}}}]{Franco:2000}
\bibinfo{author}{\bibfnamefont{L.~M.} \bibnamefont{{Franco}}},
  \bibinfo{author}{\bibfnamefont{B.}~\bibnamefont{{Link}}}, \bibnamefont{and}
  \bibinfo{author}{\bibfnamefont{R.~I.} \bibnamefont{{Epstein}}},
  \bibinfo{journal}{\apj} \textbf{\bibinfo{volume}{543}}, \bibinfo{pages}{987}
  (\bibinfo{year}{2000}), \eprint{astro-ph/9911105}.

\bibitem[{\citenamefont{{van Riper} et~al.}(1991)\citenamefont{{van Riper},
  {Epstein}, and {Miller}}}]{vanRiper:1991}
\bibinfo{author}{\bibfnamefont{K.~A.} \bibnamefont{{van Riper}}},
  \bibinfo{author}{\bibfnamefont{R.~I.} \bibnamefont{{Epstein}}},
  \bibnamefont{and} \bibinfo{author}{\bibfnamefont{G.~S.}
  \bibnamefont{{Miller}}}, \bibinfo{journal}{Astrophys. J. Lett.}
  \textbf{\bibinfo{volume}{381}}, \bibinfo{pages}{L47} (\bibinfo{year}{1991}).

\bibitem[{\citenamefont{{Dodson} et~al.}(2002)\citenamefont{{Dodson}, {Lewis},
  and {McCulloch}}}]{Dodson:2002}
\bibinfo{author}{\bibfnamefont{R.}~\bibnamefont{{Dodson}}},
  \bibinfo{author}{\bibfnamefont{D.~R.} \bibnamefont{{Lewis}}},
  \bibnamefont{and} \bibinfo{author}{\bibfnamefont{P.~M.}
  \bibnamefont{{McCulloch}}}, in \emph{\bibinfo{booktitle}{ASP Conf. Ser. 271:
  Neutron Stars in Supernova Remnants}} (\bibinfo{year}{2002}), p.
  \bibinfo{pages}{357}.

\bibitem[{\citenamefont{{Lyne} et~al.}(1993)\citenamefont{{Lyne}, {Pritchard},
  and {Graham-Smith}}}]{Lyne:1993}
\bibinfo{author}{\bibfnamefont{A.~G.} \bibnamefont{{Lyne}}},
  \bibinfo{author}{\bibfnamefont{R.~S.} \bibnamefont{{Pritchard}}},
  \bibnamefont{and}
  \bibinfo{author}{\bibfnamefont{F.}~\bibnamefont{{Graham-Smith}}},
  \bibinfo{journal}{Mon. Not. R. Astron. Soc.} \textbf{\bibinfo{volume}{265}},
  \bibinfo{pages}{1003} (\bibinfo{year}{1993}).

\bibitem[{\citenamefont{{Hurley} et~al.}(2005)\citenamefont{{Hurley}, {Boggs},
  {Smith}, {Duncan}, {Lin}, {Zoglauer}, {Krucker}, {Hurford}, {Hudson},
  {Wigger} et~al.}}]{Hurley:2005}
\bibinfo{author}{\bibfnamefont{K.}~\bibnamefont{{Hurley}}},
  \bibinfo{author}{\bibfnamefont{S.~E.} \bibnamefont{{Boggs}}},
  \bibinfo{author}{\bibfnamefont{D.~M.} \bibnamefont{{Smith}}},
  \bibinfo{author}{\bibfnamefont{R.~C.} \bibnamefont{{Duncan}}},
  \bibinfo{author}{\bibfnamefont{R.}~\bibnamefont{{Lin}}},
  \bibinfo{author}{\bibfnamefont{A.}~\bibnamefont{{Zoglauer}}},
  \bibinfo{author}{\bibfnamefont{S.}~\bibnamefont{{Krucker}}},
  \bibinfo{author}{\bibfnamefont{G.}~\bibnamefont{{Hurford}}},
  \bibinfo{author}{\bibfnamefont{H.}~\bibnamefont{{Hudson}}},
  \bibinfo{author}{\bibfnamefont{C.}~\bibnamefont{{Wigger}}},
  \bibnamefont{et~al.}, \bibinfo{journal}{Nature}
  \textbf{\bibinfo{volume}{434}}, \bibinfo{pages}{1098} (\bibinfo{year}{2005}),
  \eprint{astro-ph/0502329}.

\bibitem[{LAL()}]{LAL}
\emph{\bibinfo{title}{{LIGO Algorithm Library}}},
  \bibinfo{note}{\url{www.lsc-group.phys.uwm.edu/daswg/projects/lal.html}}.

\bibitem[{\citenamefont{{Goggin} and {the LIGO Scientific
  Collaboration}}(2006)}]{Goggin:2006}
\bibinfo{author}{\bibfnamefont{L.~M.} \bibnamefont{{Goggin}}} \bibnamefont{and}
  \bibinfo{author}{\bibnamefont{{the LIGO Scientific Collaboration}}},
  \bibinfo{journal}{Classical and Quantum Gravity}
  \textbf{\bibinfo{volume}{23}}, \bibinfo{pages}{709} (\bibinfo{year}{2006}).

\bibitem[{\citenamefont{{Mukherjee (on behalf of the LIGO Science
  Collaboration)}}(2006)}]{soma:glitches}
\bibinfo{author}{\bibfnamefont{S.}~\bibnamefont{{Mukherjee (on behalf of the
  LIGO Science Collaboration)}}}, \bibinfo{journal}{Classical and Quantum
  Gravity} \textbf{\bibinfo{volume}{23}}, \bibinfo{pages}{S661}
  (\bibinfo{year}{2006}),
  \urlprefix\url{http://stacks.iop.org/0264-9381/23/S661}.

\bibitem[{\citenamefont{{Ajith} et~al.}(2006)\citenamefont{{Ajith}, {Hewitson},
  {Smith}, and {Strain}}}]{ajith:2006}
\bibinfo{author}{\bibfnamefont{P.}~\bibnamefont{{Ajith}}},
  \bibinfo{author}{\bibfnamefont{M.}~\bibnamefont{{Hewitson}}},
  \bibinfo{author}{\bibfnamefont{J.~R.} \bibnamefont{{Smith}}},
  \bibnamefont{and} \bibinfo{author}{\bibfnamefont{K.~A.}
  \bibnamefont{{Strain}}}, \bibinfo{journal}{Classical and Quantum Gravity}
  \textbf{\bibinfo{volume}{23}}, \bibinfo{pages}{5825} (\bibinfo{year}{2006}),
  \eprint{gr-qc/0605079}.

\bibitem[{\citenamefont{{Fritschel}}(2003)}]{Fritschel:2003}
\bibinfo{author}{\bibfnamefont{P.}~\bibnamefont{{Fritschel}}}, in
  \emph{\bibinfo{booktitle}{Gravitational-Wave Detection. Edited by Cruise,
  Mike; Saulson, Peter. Proceedings of the SPIE, Volume 4856, pp. 282-291
  (2003)}}, edited by
  \bibinfo{editor}{\bibfnamefont{M.}~\bibnamefont{{Cruise}}} \bibnamefont{and}
  \bibinfo{editor}{\bibfnamefont{P.}~\bibnamefont{{Saulson}}}
  (\bibinfo{year}{2003}), pp. \bibinfo{pages}{282--291}.

\bibitem[{\citenamefont{{Willke} et~al.}(2006)\citenamefont{{Willke}, {Ajith},
  {Allen}, {Aufmuth}, {Aulbert}, {Babak}, {Balasubramanian}, {Barr},
  {Berukoff}, {Bunkowski} et~al.}}]{GEOHF}
\bibinfo{author}{\bibfnamefont{B.}~\bibnamefont{{Willke}}},
  \bibinfo{author}{\bibfnamefont{P.}~\bibnamefont{{Ajith}}},
  \bibinfo{author}{\bibfnamefont{B.}~\bibnamefont{{Allen}}},
  \bibinfo{author}{\bibfnamefont{P.}~\bibnamefont{{Aufmuth}}},
  \bibinfo{author}{\bibfnamefont{C.}~\bibnamefont{{Aulbert}}},
  \bibinfo{author}{\bibfnamefont{S.}~\bibnamefont{{Babak}}},
  \bibinfo{author}{\bibfnamefont{R.}~\bibnamefont{{Balasubramanian}}},
  \bibinfo{author}{\bibfnamefont{B.~W.} \bibnamefont{{Barr}}},
  \bibinfo{author}{\bibfnamefont{S.}~\bibnamefont{{Berukoff}}},
  \bibinfo{author}{\bibfnamefont{A.}~\bibnamefont{{Bunkowski}}},
  \bibnamefont{et~al.}, \bibinfo{journal}{Classical and Quantum Gravity}
  \textbf{\bibinfo{volume}{23}}, \bibinfo{pages}{207} (\bibinfo{year}{2006}).

\bibitem[{\citenamefont{{Ferrari} et~al.}(2003)\citenamefont{{Ferrari},
  {Miniutti}, and {Pons}}}]{SNe_ringdowns}
\bibinfo{author}{\bibfnamefont{V.}~\bibnamefont{{Ferrari}}},
  \bibinfo{author}{\bibfnamefont{G.}~\bibnamefont{{Miniutti}}},
  \bibnamefont{and} \bibinfo{author}{\bibfnamefont{J.~A.}
  \bibnamefont{{Pons}}}, \bibinfo{journal}{Mon. Not. R. Astron. Soc.}
  \textbf{\bibinfo{volume}{342}}, \bibinfo{pages}{629} (\bibinfo{year}{2003}),
  \eprint{astro-ph/0210581}.

\bibitem[{\citenamefont{{Dimmelmeier} et~al.}(2002)\citenamefont{{Dimmelmeier},
  {Font}, and {M{\"u}ller}}}]{dimmelmeier}
\bibinfo{author}{\bibfnamefont{H.}~\bibnamefont{{Dimmelmeier}}},
  \bibinfo{author}{\bibfnamefont{J.~A.} \bibnamefont{{Font}}},
  \bibnamefont{and}
  \bibinfo{author}{\bibfnamefont{E.}~\bibnamefont{{M{\"u}ller}}},
  \bibinfo{journal}{Astron. Astrophys.} \textbf{\bibinfo{volume}{393}},
  \bibinfo{pages}{523} (\bibinfo{year}{2002}), \eprint{astro-ph/0204289}.

\end{thebibliography}
\end{document}